\documentclass[acmsmall]{acmart}

\usepackage{graphicx}
\usepackage{hyperref}
\usepackage{amsmath}
\usepackage{enumerate}
\usepackage{xspace}
\usepackage{algorithm}
\usepackage{enumitem}
\usepackage{enumerate}
\usepackage{subfigure}
\usepackage[noend]{algpseudocode}
\usepackage{threeparttable}
\usepackage{xcolor}
\usepackage{tcolorbox}
\usepackage{multirow}
\usepackage{caption}

\newtheorem{theorem}{Theorem}
\newtheorem{mycorollary}{Corollary}{\bfseries}{\rmfamily}

\newcommand{\specialcell}[2][c]{
	\begin{tabular}[#1]{@{}c@{}}#2\end{tabular}}

\newcommand{\romannum}[1]{\uppercase\expandafter{\romannumeral #1\relax}}

\newcommand{\algrule}[1][.2pt]{\par\vskip.5\baselineskip\hrule height #1\par\vskip.5\baselineskip}
\mathchardef\mhyphen="2D
\newcommand{\infhors}{\ensuremath {\texttt{LiteQS} {\xspace} }}
\newcommand{\infhorskg}{\ensuremath {\texttt{\infhors.Kg}{\xspace}}}
\newcommand{\infhorssig}{\ensuremath {\texttt{\infhors.Sig}{\xspace}}}
\newcommand{\infhorsver}{\ensuremath {\texttt{\infhors.Ver}{\xspace}}}
\newcommand{\infhorspkconstr}{\ensuremath {\texttt{\infhors.PKConstr}{\xspace}}}

\newcommand{\DM}{\ensuremath {\texttt{DM}{\xspace}}}
\newcommand{\fhe}{\ensuremath {\texttt{FHE}{\xspace}}}

\newcommand{\eval}{\ensuremath {\texttt{Eval}{\xspace}}}
\newcommand{\enc}{\ensuremath {\texttt{Enc}{\xspace}}}
\newcommand{\dec}{\ensuremath {\texttt{Dec}{\xspace}}}

\newcommand{\fhekg}{\ensuremath {\texttt{FHE.Kg}{\xspace}}}
\newcommand{\fheenc}{\ensuremath {\texttt{FHE.Enc}{\xspace}}}
\newcommand{\fhedec}{\ensuremath {\texttt{FHE.Dec}{\xspace}}}
\newcommand{\fheeval}{\ensuremath {\texttt{FHE.Eval}{\xspace}}}
\newcommand{\eucma}{\ensuremath {\texttt{EU}\mbox{-}\texttt{CMA}{\xspace}}}

\newcommand{\owf}{\ensuremath {\texttt{OWF}{\xspace}}}
\newcommand{\hors}{\ensuremath {\texttt{HORS}{\xspace}}}

\newcommand{\pkosgnkg}{\ensuremath {\texttt{PKO}\mbox{-}\texttt{SGN.Kg}}{\xspace}}
\newcommand{\pkosgnsig}{\ensuremath {\texttt{PKO}\mbox{-}\texttt{SGN.Sig}}{\xspace}}
\newcommand{\pkosgnpkconstr}{\ensuremath {\texttt{PKO}\mbox{-}\texttt{SGN.PKConstr}}{\xspace}}
\newcommand{\pkosgnver}{\ensuremath {\texttt{PKO}\mbox{-}\texttt{SGN.Ver}}{\xspace}}
\newcommand{\pkosgn}{\ensuremath {\texttt{PKO}\mbox{-}\texttt{SGN}}{\xspace}}
\newcommand{\mpk}{\ensuremath {\mathit{MPK}}{\xspace}}
\newcommand{\csk}{\ensuremath {\mathit{csk}}{\xspace}}

\newcommand{\advsaleh}{\ensuremath {\mathit{Adv}_{\mathit{\pkosgn}}^{\eucma}}{\xspace}}

\newcommand{\adveucmaINFHORS}{\ensuremath {\mathit{Adv}_{\mathit{\infhors}}^{\eucma}}{\xspace}}
\newcommand{\adveucmaHORS}{\ensuremath {\mathit{Adv}_{\mathit{\horsss}}^{\eucma}}{\xspace}}

\newcommand{\xmss}{\ensuremath {\text{XMSS}^{\text{MT}}{\xspace}}}
\newcommand{\himq}{\ensuremath {\text{HiMQ-3}^{\text{Big}}{\xspace}}}

\newcommand{\A}{$\mathcal{A}$~}
\newcommand{\F}{$\mathcal{F}$~}

\newcommand{\Ra}{\ensuremath \stackrel{\$}{\leftarrow}{\xspace}}

\newcommand{\xor}{\oplus}
\newcommand{\as}{\ensuremath {\leftarrow}{\xspace}}

\newcommand{\lm}{\ensuremath {\mathcal{LM}}{\xspace}}
\newcommand{\ls}{\ensuremath {\mathcal{LS}}{\xspace}}

\newcommand{\msk}{\ensuremath {\mathit{msk}}{\xspace}}
\newcommand{\sk}{\ensuremath {\mathit{sk}}{\xspace}}

\newcommand{\pk}{\ensuremath {\mathit{PK}}{\xspace}}

\newcommand{\prf}{\ensuremath {\texttt{PRF}{\xspace}}}
\newcommand{\cmp}{\ensuremath {\texttt{CMP}{\xspace}}}

\newcommand{\kg}{\ensuremath {\texttt{Kg}}{\xspace}}
\newcommand{\ssig}{\ensuremath {\texttt{Sig}}{\xspace}}
\newcommand{\ver}{\ensuremath {\texttt{Ver}}{\xspace}}
\newcommand{\ppkconstr}{\ensuremath {\texttt{PKConstr}}{\xspace}}

\newcommand{\horsss}{\ensuremath {\texttt{HORS}{\xspace}}}
\newcommand{\horskg}{\ensuremath {\texttt{HORS.Kg}}{\xspace}}
\newcommand{\horssig}{\ensuremath {\texttt{HORS.Sig}}{\xspace}}
\newcommand{\horsver}{\ensuremath {\texttt{HORS.Ver}}{\xspace}}

\AtBeginDocument{%
}

\setcopyright{acmcopyright}

\acmJournal{TIOT}

\begin{document}

\title{LiteQSign: Lightweight and Quantum-Safe Signatures for Heterogeneous IoT Applications}

\author{Attila A. Yavuz}
\email{attilaayavuz@usf.edu}
\author{Saleh Darzi}
\email{salehdarzi@usf.edu}
\author{Saif E. Nouma}
\email{saifeddinenouma@usf.edu}
\affiliation{%
		\institution{University of South Florida}
		\streetaddress{3720 Spectrum Blvdve, Interdisciplinary Research Building (IDR)-400}
		\city{Tampa}
		\state{Florida}
		\country{USA}
		\postcode{33612}
	}






\begin{abstract}	
The rapid proliferation of resource-constrained IoT devices across sectors like healthcare, industrial automation, and finance introduces major security challenges. Traditional digital signatures, though foundational for authentication, are often infeasible for low-end devices with limited computational, memory, and energy resources. Also, the rise of quantum computing necessitates post-quantum (PQ) secure alternatives. However, NIST-standardized PQ signatures impose substantial overhead, limiting their practicality in energy-sensitive applications such as wearables, where signer-side efficiency is critical. 
To address these challenges, we present LightQSign (\infhors), a novel lightweight PQ signature that achieves near-optimal signature generation efficiency with only a small, constant number of hash operations per signing. Its core innovation enables verifiers to obtain one-time hash-based public keys without interacting with signers or third parties through secure computation. We formally prove the security of \infhors~in the random oracle model and evaluate its performance on commodity hardware and a resource-constrained 8-bit AtMega128A1 microcontroller. Experimental results show that \infhors~outperforms NIST PQC standards with lower computational overhead, minimal memory usage, and compact signatures. On an 8-bit microcontroller, it achieves up to $1.5\mbox{--}24\times$ higher energy efficiency and $1.7\mbox{--}22\times$ shorter signatures than PQ counterparts, and $56\mbox{--}76\times$ better energy efficiency than conventional standards—enabling longer device lifespans and scalable, quantum-resilient authentication.

\end{abstract}




\maketitle

\section{Introduction} \label{sec:Introduction}

The proliferation of resource-constrained IoT devices, coupled with their extensive integration into critical domains such as healthcare, industrial automation, and financial services, introduces significant security challenges. While these devices deliver substantial utility and advancements, their operation in challenging environments and inherent limitations make them the most vulnerable link in the security chain. With billions of IoT devices interconnected across networks, many function in environments that process sensitive or personal data, including healthcare records \cite{ahad2020technologies}, military communications \cite{pradhan2020security}, and security logs \cite{behnia2021towards,camara2020access}, while executing essential operations such as unlocking smart doors, regulating industrial machinery \cite{nguyen20216g}, and administering medical treatments \cite{joung2013development}. For instance, compromised authentication undermines the integrity of data from implantable devices, rendering them ineffective and potentially endangering patients, such as failing to correct a slow heartbeat in time \cite{sametinger2015security}.  Therefore, ensuring device authenticity and data integrity is essential for protecting networks against malicious entities, preserving reliable data transmission, and preventing unauthorized access to critical control systems. Moreover, IoT devices rely on periodic updates to mitigate vulnerabilities and enhance functionality; verifying the authenticity and integrity of these updates is vital to prevent unauthorized or malicious software installation and to maintain overall system security \cite{mudgerikar2021iot, behnia2021towards, kim2020resilient}.



A ``\textit{digital signature}'' serves as the cornerstone cryptographic solution for ensuring authentication and data integrity while providing essential features such as non-repudiation, public verifiability, and scalability \cite{simsek2024authentication, Yavuz:HASES:ICC2023}. These properties enable IoT systems to operate securely and efficiently, mitigating risks of unauthorized access, data breaches, device manipulation, and network disruptions.  Given the diverse operational constraints of IoT applications—particularly in computation and energy efficiency—authentication mechanisms must be tailored accordingly. Here, we focus on IoT-enabled applications that prioritize lightweight signing due to resource constraints, while slower verification is acceptable on backend systems. Numerous real-world IoT use cases align with this paradigm. Examples include environmental and industrial sensors that periodically transmit data, such as temperature or air quality, to central servers, while smart utility meters (e.g., energy, water, gas) frequently report usage data \cite{sisodia2024two, goudarzi2022survey, ji2020authenticating}. Wireless sensor networks with verifier-side delay tolerance \cite{yavuz2012self} and wearable health devices continuously collect biometric metrics, forwarding them to paired devices or the cloud \cite{adamson2009pathophysiology,zile2008transition, li2023mmhsv}. Remote IoT nodes, including wildlife trackers and agricultural sensors \cite{vangala2023security}, operate in power-constrained environments, transmitting critical updates. RFID/NFC-based asset tracking systems enable identification with resourceful verifiers \cite{rieback2005rfid, shi2021wifi}, while smart city sensors for traffic, pollution, and waste management autonomously collect and transmit data on low power, prioritizing lightweight signing and deferred verification \cite{zhang2020safecity}.


Therefore, designing practical authentication solutions for these heterogeneous IoT-enabled applications necessitates addressing the following key considerations:

{\em (I)} \underline{\textit{Post-Quantum Security}}:
The rise of quantum computing threatens traditional digital signatures based on number-theoretic assumptions (e.g., factorization, discrete logarithm) \cite{shor1999polynomial,aumasson2017impact, darzi2023envisioning}, necessitating long-term security solutions with PQ guarantees for IoT applications handling sensitive data and critical operations. Many IoT devices have long lifespans and cannot easily transition to new cryptographic frameworks once deployed, making immediate PQ adoption essential \cite{cheng2017securing}. For instance, updating security measures on heart implants requires surgery, and medical audit logs must remain verifiable for decades \cite{adamson2009pathophysiology}, while upgrading mass-produced smart meters at scale is costly and impractical \cite{adeli2023post}. Moreover, resource constraints such as limited processing power, memory, and battery life (e.g., 8-bit microcontrollers) further exacerbate PQ deployment. Addressing these challenges is critical for securing IoT-enabled environments, ensuring long-term resilience against emerging quantum threats \cite{nouma2024trustworthy}.

{\em (II)} \underline{\textit{Computationally Efficient Signature Generation with Minimal Energy Usage:}} The primary design consideration is lightweight signature generation with signer-efficient operations and minimal energy consumption. This is especially critical for battery-powered embedded IoT devices~\cite{camara2015security}, where energy efficiency directly affects system longevity. For instance, in wearables, prolonged battery life enhances usability \cite{camara2020access}, while for implantable devices, it directly impacts patient well-being, as replacements may require surgical intervention \cite{camara2015security}. Hence, the underlying cryptographic mechanism must minimize energy consumption to extend device longevity. However, even standard digital signatures based on conventional security (e.g., Elliptic Curve Cryptography) have been shown to degrade battery life in low-end IoT devices \cite{martinez2023comprehensive, nouma2023practical}. This challenge is further intensified when PQ security is considered \cite{shim2020high, cheng2017securing}.

{\em (III)} \underline{\textit{Minimal Cryptographic Memory and Bandwidth Usage}}: {\em (i)} Embedded IoT platforms, such as the 8-bit ATxmega128A1 MCU with only $128~KB$ of flash memory\footnote{\url{https://www.microchip.com/en-us/product/atxmega128a1}}, demand strict memory efficiency. In devices with severe memory and bandwidth constraints (e.g., heart implants, wearables), NIST-standardized signatures like ML-DSA and SLH-DSA—requiring approximately $6.2~\textit{KB}$ and $71.9~\textit{KB}$ for combined signature, private, and public keys—are unsuitable for deployment. Thus, lightweight signatures must feature compact keys and minimal memory expansion during computation.  
(ii) Code size is another constraint, where schemes relying on simple primitives (e.g., hashing) over computationally intensive operations (e.g., EC scalar multiplication \cite{costello2016schnorrq}, sampling \cite{dang2024module}) offer reduced energy usage and implementation overhead.    
(iii) Larger signature sizes increase memory and bandwidth demands and accelerate battery drain during transmission \cite{sehgal2012management}. These limitations make large PQ-secure signatures particularly challenging for resource-constrained IoT devices.




{\em (IV)} \underline{\textit{Security Assumptions and Robustness}} 
{\em (i)} Minimal Security Assumptions: While some schemes improve efficiency by assuming semi-honest, non-colluding servers or secure enclaves \cite{nouma2024trustworthy, behnia2021towards}—the security-sensitive nature of IoT-enabled applications necessitates avoiding such dependencies. Eliminating extra architectural assumptions strengthens long-term trust, not only through PQ cryptographic guarantees but also by enhancing resilience against emerging threats.  
{\em (ii)} Robustness Against Side-Channel Attacks: PQ signatures involve Gaussian sampling \cite{fouque2018falcon}, rejection sampling \cite{dang2024module}, and complex arithmetic \cite{ducas2013lattice, shim2020high}, making them vulnerable to side-channel attacks \cite{karabulut2021falcon, tibouchi2021one}. These risks are amplified in embedded architectures due to the difficulty of implementing countermeasures \cite{camara2020access}. Additionally, low-end IoT devices often generate low-quality random numbers, exposing them to cryptographic vulnerabilities \cite{rfc6979}. Addressing these challenges is essential for securing IoT systems. 
{\em (iii)} Scalability and Interoperability: Ensuring seamless communication among diverse devices in heterogeneous IoT ecosystems requires an authentication scheme that efficiently scales across millions of interconnected, resource-limited devices. Given the challenges of key management in large-scale IoT networks with constrained connectivity, maintaining constant-size public keys and lightweight mechanisms for key distribution, renewal, and revocation is essential.

\subsection{Related Work and Limitations of the State-of-the-Art} \label{subsec:Relatedwork}

This section examines state-of-the-art signatures that address key considerations for IoT-enabled applications, with a focus on schemes offering PQ security, computationally efficient signing, and compact public key and signature sizes. Given the extensive range of proposed signatures, we first provide a brief overview of prominent conventional signatures before shifting to PQ alternatives, including both standardized schemes and those optimized for signing efficiency.

{\em Conventional lightweight signatures}: These schemes offer efficient signature generation, compact key sizes, and additional security guarantees \cite{nouma2023practical, verma2019cb, chen2022secure}. Among conventional signatures, EC Schnorr-based schemes \cite{costello2016schnorrq} are particularly efficient compared to pairing-based \cite{liu2023idenmultisig} and factorization-based \cite{zhang2020security} alternatives. For instance, a recent EC-based scheme \cite{nouma2023practical} enhances signer efficiency and supports single-signer signature aggregation, while Chen et al. \cite{chen2022secure} integrate confidentiality with authentication. However, despite their advantages, none of these ECC-based schemes or similar conventional signatures provide PQ security.

{\em Standard PQ signatures}: NIST has recently standardized Module-Lattice-Based Digital Signature Standard (ML-DSA FIPS 204 \cite{dang2024module}) as a lattice-based and Stateless Hash-Based Digital Signature
Standard (SLH-DSA FIPS 205 \cite{cooper2024stateless}) as a hash-based digital signature standard \cite{nist-pqc}. The following discusses these domains in detail.

Known for their minimal intractability assumptions, hash-based signatures provide strong security guarantees. The stateless SLH-DSA \cite{cooper2024stateless}, derived from a variant of the one-time signature scheme HORS \cite{reyzin2002better}, employs a hyper-tree structure for multiple-time signatures. While SLH-DSA ensures PQ security with strong assumptions, its signing time and signature size are orders of magnitude slower and larger than ECDSA, making it unsuitable for resource-constrained IoT devices. Similarly, stateful hash-based signatures (e.g., RFC-standardized XMSS-MT \cite{hulsing2017optimal} and LMS \cite{rfc8554}), built on HORS variants like W-OTS \cite{merkle1989certified}, offer comparable security and forward security. However, their state management requirements, high computational cost, and memory demands render them impractical for low-end IoT devices.

Based on module lattice problems (e.g., LWE \cite{dang2024module}), lattice-based signatures offer a balanced trade-off between signing and verification efficiency. NIST-selected schemes, ML-DSA \cite{dang2024module} and Falcon \cite{fouque2018falcon}, achieve smaller signatures and faster signing than SLH-DSA. However, they remain unsuitable for resource-constrained IoT devices due to their computational complexity and larger signature sizes compared to conventional signatures. Additionally, techniques such as Gaussian and rejection sampling introduce vulnerabilities to side-channel attacks \cite{karabulut2021falcon}. To date, no open-source lattice-based signature implementation is optimized for highly constrained devices like 8-bit microcontrollers, except for BLISS \cite{ducas2013lattice}, which was not selected as a NIST PQC standard \cite{yavuz2023beyond} and is susceptible to side-channel attacks \cite{tibouchi2021one}.


{\em Additional PQ Signatures for Standardization}: To diversify PQ signature standards, NIST launched a competition alongside its standardized schemes, emphasizing fast verification and compact signatures. Currently in its second round, it includes submissions across various PQC categories, such as code-based \cite{baldi2024zero}, multivariate-based \cite{beullens2021mayo}, and symmetric-based \cite{kim2023aim} signatures. For instance, Shim et al. \cite{shim2020high} improves upon NIST PQC standards with efficient signing and smaller signatures but suffers from large private key and code sizes, making it impractical for IoT devices. Its $12.6 \textit{KB}$ private key is an order of magnitude larger than ECDSA, and its implementation occupies $62.6\%$ of the total flash memory on an ATxmega128A1 MCU, imposing significant memory overhead. Despite PQC advancements, computational and memory constraints remain major obstacles for low-end IoT devices, particularly those operating on 8-bit MCUs \cite{yavuz2023beyond}. Also, many multivariate signatures not only demand extensive memory and stack resources but are also susceptible to polynomial-time attacks, compromising their unforgeability \cite{hashimoto2011general}.

{\em Lightweight PQ signatures with Additional Assumptions}: These schemes achieve highly efficient signature generation but rely on additional assumptions \cite{behnia2021towards, nouma2024trustworthy, ouyang2021scb}. For example, ANT \cite{behnia2021towards}, a lattice-based signature, delegates costly commitment generation to distributed, non-colluding, semi-honest servers \cite{sedghighadikolaei2023comprehensive}. However, verifiers must interact with these servers before verification, introducing potential network delays and outage risks. Another approach leverages Trusted Execution Environment (TEE)-assisted signatures, offloading computational overhead to TEE-enabled servers \cite{ouyang2021scb, Yavuz:HASES:ICC2023, nouma2024trustworthy}. For instance, HASES \cite{Yavuz:HASES:ICC2023} and its extension \cite{nouma2024trustworthy}, derived from the one-time HORS \cite{reyzin2002better}, use a single TEE-enabled cloud server to issue one-time public keys. However, reliance on a centralized TEE server introduces key escrow risks and a single point of trust. Due to these additional assumptions, such signatures may not be ideal for IoT-enabled applications that require adherence to traditional public key settings.

Given the limitations of existing signatures and the gap in achieving all desirable properties for IoT applications, there is a critical need for efficient PQ signatures that balance performance and security while enabling signer-optimal generation and deferred verification. This work explores the following key research questions:  
\textit{(i) Can an efficient PQ signature scheme be designed with optimal signature generation while meeting IoT constraints on memory, processing, and bandwidth?} 
\textit{(ii) Is it possible to achieve energy-efficient signing without introducing unconventional or risky assumptions for verifiers?} 
\textit{(iii) Can these requirements be met in a scalable multi-user setting for IoT networks?}


\vspace{-4mm}
\subsection{Our Contribution}
{\em We propose \ensuremath{\texttt{LiteQSign}}, a novel lightweight PQ signature that enables verifiers to derive one-time public keys independently, eliminating the need for interaction with signers or third parties. By extending the one-time $\hors$ scheme into a multiple-time signature, $\infhors$ allows verifiers to extract one-time keys from a master public key via encrypted pseudo-random function evaluations using fully homomorphic encryption (FHE). To the best of our knowledge, this is the first $\hors$-type hash-based scheme that lifts message-signing constraints without relying on semi-honest servers (ANT \cite{behnia2021towards}), trusted hardware (HASES \cite{Yavuz:HASES:ICC2023}), or the high computational and storage costs of XMSS \cite{hulsing2013optimal} and SLH-DSA \cite{cooper2024stateless}.} Our design offers several desirable properties, as follows:

$\bullet$ \textbf{\textit{Signer-Optimal Computation with Energy Efficiency}}: Our scheme makes a single call to the one-time $\hors$ signature scheme while maintaining identical signature sizes, achieving \textit{optimal efficiency} relative to $\hors$. Thus, $\infhors$ requires only a small, constant number (e.g., $16$) of Pseudo-Random Function (PRF) calls per signing. As shown in Table \ref{tab:embeddedperformance}, it achieves ${19.5\times}$ and ${1130\times}$ faster signature generation compared to NIST PQC standards ML-DSA-\romannum{2} and Falcon-512, respectively. Also, it surpasses even the most efficient ECC-based conventional signatures. As detailed in Section \ref{sec:performance}, this computational efficiency translates directly into significant energy savings.

$\bullet$ {\textbf{\textit{Minimal Memory Requirements and Bandwidth Efficiency}}: 
The private key of $\infhors$ consists solely of a single random seed (e.g., 128-bit), and it transmits only one $\hors$ signature per message (i.e., $256$ bytes), making it the most compact among PQ counterparts (see Table \ref{tab:embeddedperformance}). Unlike multivariate- \cite{shim2020high} and lattice-based \cite{ducas2013lattice} schemes, $\infhors$ has a minimal code size for signature generation. It requires only a few PRF calls and a single hash call, avoiding costly operations like EC scalar multiplication \cite{verma2019cb} and sampling \cite{tibouchi2021one}. Moreover, its use of AES-128 as the PRF and SHA-256 as the hash function ensures standard compliance, facilitating an efficient transition to PQC \cite{nouma2024trustworthy}.

$\bullet$ \textbf{\textit{Advanced Security Features and Robustness}}: 
{\em (i)} $\infhors$ adheres to the standard public key model, avoiding unconventional assumptions such as non-colluding or trusted key distribution servers \cite{behnia2021towards}, which introduce architectural risks. {\em (ii)} Unlike lattice-based schemes prone to side-channel and timing attacks due to Gaussian and rejection sampling \cite{karabulut2021falcon}, $\infhors$ relies solely on symmetric cryptographic primitives, eliminating these vulnerabilities. Also, its deterministic signature generation prevents weaknesses from poor random number generators, a common issue in resource-limited IoT devices.

$\bullet$ \textbf{\textit{Compact Multi-User Storage and Online/Offline Verification}}: 
$\infhors$ offers a scalable solution by allowing verifiers to derive one-time public keys from a constant-size master public key, removing the need for storing per-user keys or certificates—even for large-scale deployments (e.g., $2^{20}$ users). Public keys can be precomputed or generated on-demand before verification. Though key construction involves encrypted function evaluations, it can be performed independently by the verifier or offloaded to a resourceful cloud server, significantly minimizing its practical impact.

$\bullet$ \underline{\textbf{\textit{Limitations and Potential Use-Case}}}: 
$\infhors$ is particularly suited for lightweight authentication in resource-limited IoT applications, offering near-optimal signer efficiency with delay-tolerant, resourceful verifiers. $\infhors$ prioritizes high PQ security and robustness, excelling in IoT scenarios where the signer is non-interactive and must operate with maximum efficiency (e.g., medical wearables, trackers), while verification can tolerate some delay and storage overhead. In such use cases, device efficiency and battery longevity are paramount. As detailed in Section \ref{sec:securitymodel} (Figure \ref{fig:systemmodel}), $\infhors$ is well-suited for heterogeneous IoT environments \cite{yavuz2013eta}.

\section{Preliminaries} \label{sec:preliminaries}
This section presents the notation and acronyms in Table~\ref{tab:notations}, followed by an overview of the cryptographic primitives that form the foundation of our scheme.
\vspace{2pt}

\begin{table}[ht!]
	\centering
	\caption{Acronyms and notations} 
	\label{tab:notations}
	\resizebox{0.75\textwidth}{!}{
		\begin{tabular}{|l|l|}
			\hline
			\textbf{Notation} & \textbf{Description} \\ \hline \hline
			
			PQC & Post-Quantum Cryptography \\ \hline
			ECC & Elliptic Curve Cryptography \\ \hline
			FHE & Fully Homomorphic Encryption \\ \hline
			HORS & Hash to Obtain Random Subset \\ \hline
			PKO-SGN & Public Key Outsourced Signature \\ \hline
			EU-CMA & Existential Unforgeability against Chosen Message Attack \\ \hline
			IND-CPA & Indistinguishably under Chosen Plaintext Attack  \\ \hline
			ROM & Random Oracle Model \\ \hline
			PRF & Pseudo-Random Function \\ \hline 
			PPT & Probabilistic Polynomial Time \\ \hline
			OWF & One-Way Function \\ \hline
			\sk/\pk & Private/Public key \\ \hline
			\msk/\mpk & Master private/public key \\ \hline
			$ID_i / N$ & User identity (e.g., MAC address) and total number of users \\ \hline
			$j$ & Signer state \\ \hline
			$x_i$ & variable of the user $ID_i$ \\ \hline
			$x_i^j$ & variable for the user $ID_i$ with the state $j$ \\ \hline
			$x_i^{j,\ell}$ & $\ell^{\text{th}}$ element of  variable $x_i^j$ for the user $ID_i$ with the state $j$ \\ \hline
			$x \Ra \mathcal{S} / |x|$ & random selection from a set $\mathcal{S}$ and bit length of variable $x$ \\ \hline
			$|| / \xor$ & string concatenation and bitwise-XOR operation\\ \hline
			$H: \{0,1\}^* \rightarrow \{0,1\}^\kappa$ & Cryptographic hash function \\ \hline
			$f: \{0,1\}^* \rightarrow \{0,1\}^\kappa$ & One-way function \\ \hline 
			$x\as \prf(k,M)$ & accepts a key $k$ and message $M$ as input. It outputs $x$ \\ \hline
			$C \as E_{k}(m)$ & \parbox{0.4\textwidth}{Encrypts of message $m$ under the key $k$. It outputs $C$} \\ \hline
			$\{0,1\} \as \cmp(x, y)$ & \parbox{0.4\textwidth}{Equality comparison function of two (e.g., $64$-bit) numerical values $x$ and $y$} \\ \hline
		\end{tabular}
	}
\end{table}

The Davies-Meyer scheme (\DM) \cite{Preneel2005DaviesMeyerHF} is a generic and iterated cryptographic hash function based on a block cipher $E$.  In \infhors, we only rely on the one-wayness (\owf) of \DM, which is based on the IND-CPA security of the symmetric cipher $E$. 
The $\DM$ algorithm is described as follows:

\begin{definition} \label{def:DM}
	$\underline{B_n \as \DM (M,B_0)}$: Given a message $M$, a pre-defined initial value $I_{\DM} = B_0$, and a block cipher $E$ of length $k$, it splits $M$ into $n$ chunks $M=\{m_i\}_{i=1}^n$ where $n = \lceil \frac{|M|}{k} \rceil$, and computes ${B_i = E_{m_i}(B_{i-1}) \oplus m_i, \forall i = 1, 2, ..., n}$. Finally, it outputs $B_n$ as the hash output.
\end{definition}

Hash to Obtain Random Subset (\hors) \cite{reyzin2002better} is an efficient hash-based one-time signature scheme that leverages the subset-resilience property of the underlying hash function and the one-wayness of the employed pseudorandom function (\prf). \hors~is formally defined as follows:

\begin{definition} \label{def:HORS}
    \hors~consists of three core algorithms, $\hors = (\kg, \ssig, \ver)$ described as follows:
	\begin{itemize}[leftmargin=*]
		\item[-] $\underline{(\sk, \pk, I_\horsss) \as \horskg(1^\kappa)}$: Given the security parameter $\kappa$, it selects $I_\horsss \as (k, t)$, generates $t$ random $\kappa$-bit strings $\{s_i\}_{i=1}^t$, and computes $v_i \as f(s_i), \forall i=1, \ldots, t$.
		It sets $\sk \as \{s_i\}_{i=1}^t$ and $\pk \as \{v_i\}_{i=1}^t$.

		\item[-] $\underline{\sigma \as \horssig(\sk, M)}$:  Given \sk~and message $M$, it computes $h \as H(M)$. It splits $h$ into $k$ substrings $\{h_j\}_{j=1}^k$ (where $|h_j|=\log_2{t}$) and interprets them as integers $\{i_j\}_{j=1}^k$. It outputs $\sigma \as \{s_{i_j}\}_{j=1}^k$.
		
		\item[-] $\underline{b \as \horsver(\pk, M, \sigma)}$: Given \pk, $M$, and $\sigma$, it computes $\{i_j\}_{j=1}^k$ as in $\horssig(.)$. If $v_{i_j}=f(\sigma_j), \forall j=1,\ldots,k$, it returns $b=1$, otherwise $b=0$.
	\end{itemize}
\end{definition}

A Fully Homomorphic Encryption (FHE) scheme enables arbitrary computation on encrypted data without revealing the underlying plaintexts. Originating from Gentry’s seminal work in 2009 \cite{gentry2009fully}, FHE allows an entity to perform functions over ciphertexts such that the result, once decrypted, matches the output of the same function applied to the plaintexts. An \fhe~scheme is defined as follows: 

\begin{definition} \label{def:FHE}

 An FHE scheme \cite{armknecht2015guide} consists of four probabilistic polynomial-time algorithms $\fhe = (\kg, \enc, \eval, \dec)$  defined as below: 
	
	\begin{itemize}[leftmargin=*]
		\item [-] ${(\sk’, \pk’, I_{\fhe}) \as \fhekg(1^\kappa)}$: Given $\kappa$, it creates the auxiliary argument $I_{\fhe}$ and generates \fhe~private/public key pair $(\sk', \pk')$. 
		\item [-] ${C \as \fheenc(\pk’, M)}$: Given $\pk'$ and a plaintext $M$, it encrypts $M$ and returns the ciphertext $C$.
		\item [-] ${C  \as \fheeval(\pk', \mathcal{F}(\overrightarrow{c}=\{c_j\}_{j=1}^n))}$: Given $\pk'$, a function $\mathcal{F}$, and a set of input arguments $\overrightarrow{c}$, it evaluates $\mathcal{F}$ on $\overrightarrow{c}$ under encryption.
		\item [-] ${M \as \fhedec(\sk', C)}$: Given $\sk'$ and $C$, it decrypts $C$ via $\sk'$ and outputs the plaintext $M$.
	\end{itemize}
\end{definition}

For illustration, ${\fheeval(\pk', \prf(Y, x))}$ and ${\fheeval(\pk', \cmp(X_1, X_2))}$ evaluate ${\prf(y, x)}$ and ${\cmp(x_1, x_2)}$ functions under encryption, where the key $Y$ and the numerical values ($X_1, X_2$) are the encryption of $y$, $x_1$, and $x_2$ under $\pk'$ (i.e., ${Y \as \fheenc(\pk', y)}$, ${X_1 \as \fheenc(\pk', x_1)}$, ${X_2 \as \fheenc(\pk', x_2)}$), respectively. 
We choose an IND-CPA-secure FHE instantiated with the Ring Learning With Error (R-LWE) variant of the BGV cryptosystem \cite{brakerski2014leveled}. Note that these FHE instantiations also have a post-quantum security premise \cite{10.1093/nsr/nwab115}.
\vspace{2pt}

A Public Key Outsourced Signature scheme (\pkosgn)~transforms the one-time \hors~signature scheme into a multiple-time hash-based digital signature by leveraging \fhe~to provide verifiers with one-time public keys without the knowledge of the private key. Moreover, \ \pkosgn~implements the one-way function ($f$) in \hors~using the single-block-length \DM~construction.
\begin{definition} \label{def:PKOSig}
	A Public Key Outsourced Signature scheme $\pkosgn=(\kg, \ssig, \ppkconstr, \ver)$ is defined as follows:
	
	\begin{itemize}[leftmargin=*]
		\item [-] ${(\pk, \sk, I) \as \pkosgnkg(1^\kappa, \overrightarrow{ID})}$:	Given $\kappa$ and a set of users' identifiers $\overrightarrow{ID}$, it returns $\pk$ with both $\fhe$ and master public keys $\pk = \langle\pk', \mpk\rangle$, the private key $\sk=\overrightarrow{\gamma}$, and  the system-wide parameters $I \as (I_{\hors}, I{\DM}, I_\fhe)$. 
		\item [-] ${\sigma_i^j \as \pkosgnsig(\gamma_i, M_j)}$: Given the seed $\gamma_i \in \overrightarrow{\gamma}$ of $ID_i$ and  a message $M_j$, it returns the signature $\sigma_i^j$. 
		\item [-] ${cv_i^j  \as \pkosgnpkconstr(\pk, ID_i, j)}$: Given the signer $ID_i$, state $j$, and $\pk$, it constructs the required public keys under encryption $cv_i^j$ via $\fheeval(.)$. 
		\item [-] ${b \as \pkosgnver(\pk_i^j, M_j, \sigma_i^j)}$: Given $\pk_i^j$, $M_j$, and $\sigma_i^j$, it outputs $b = 1$ if $\sigma_i^j$ is valid, or $b = 0$ otherwise.
	\end{itemize}
\end{definition}

\section{System Architecture and Security Model} \label{sec:securitymodel}

\noindent \textbf{System Model}: We adopt the traditional public-key-based broadcast authentication model, tailored for diverse IoT-enabled applications while addressing key design considerations. Figure \ref{fig:systemmodel} illustrates the entities in our architecture, detailed as follows:

\begin{figure}[ht!]
	\centering
	\includegraphics[scale=0.47]{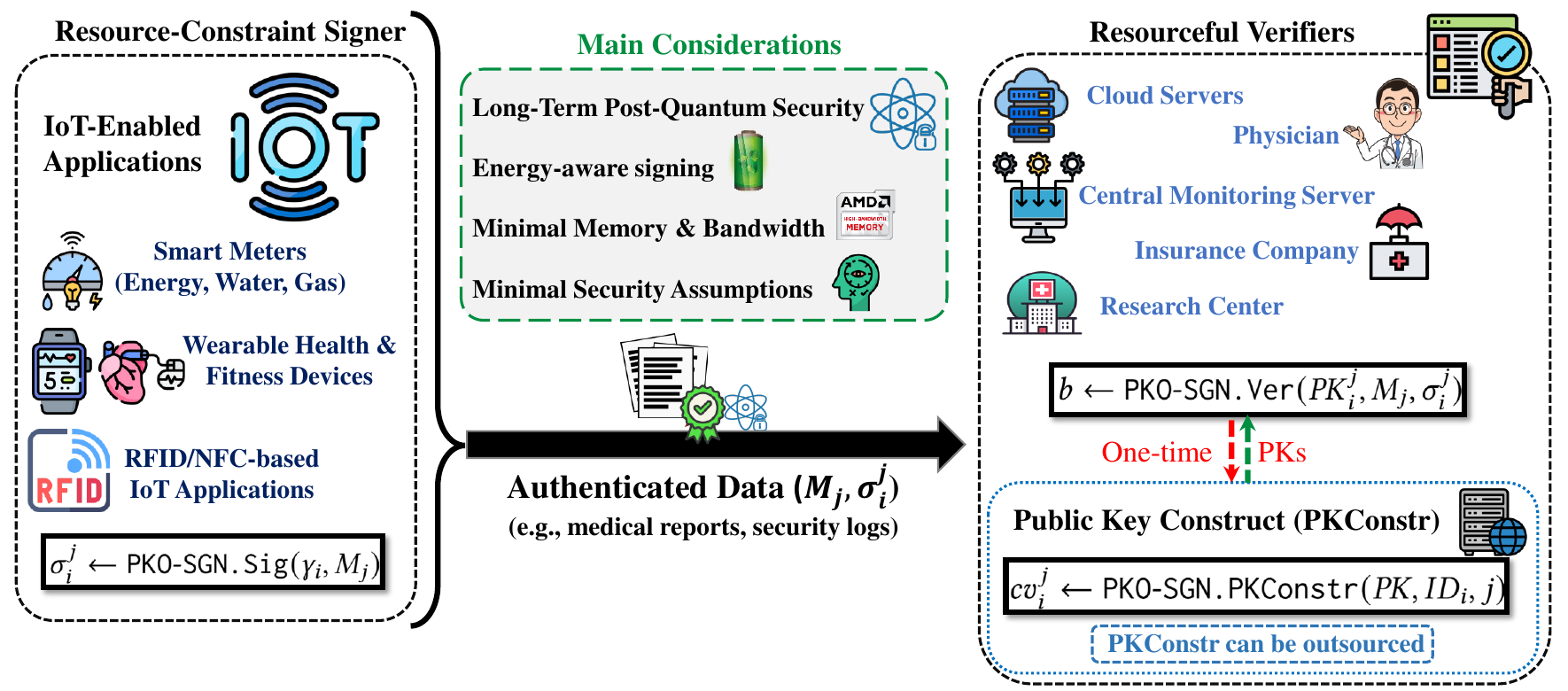}
	\caption{System model}
	\label{fig:systemmodel}
\end{figure}

$\bullet$ \textit{Signer:} A resource-constrained IoT device, such as a smart sensor, pacemaker, or implantable medical device (see Figure \ref{fig:systemmodel}), responsible for signing and broadcasting messages to verifiers. These messages often contain sensitive data (e.g., heart rate, security logs), where data integrity and source authenticity are crucial for usability. The signer prioritizes efficient, energy-aware computations with minimal memory and bandwidth usage, while ensuring long-term security.

$\bullet$ \textit{Verifier:} A resourceful entity (e.g., cloud server, physician, monitoring center) that authenticates messages from signers. It independently constructs one-time public keys from the master public key ({$\mpk$}), enabling it to derive any public key $\pk_i^j$ for any user $ID_i$ in a network of (e.g., $N=2^{20}$) signers. Additionally, we propose an optional approach where non-resourceful verifiers can outsource public key derivation to a resourceful entity (e.g., a cloud server) via $\fhe$ computations, without relying on semi-honest entities or trusted parties. \aa{further emphasis here PKConxturct is not a seperate entity, but if somebody wants, we can offload it. }

\vspace{4pt}
\noindent \textbf{Threat  and Security Model}: 
Our threat model is based on an Probabilistic Polynomial Time (PPT) adversary \A equipped with the following capabilities:

\begin{enumerate}[leftmargin=*]
    \item {\em Passive attacks:} aim to monitor and interpret the output of the signature generation interface. 
    \item {\em Active attacks:}  aim to intercept, forge, and modify messages and signatures sent from IoT devices. We assume that the adversary is equipped with a quantum computer. 
\end{enumerate}
\vspace{4pt}

We follow the standard Existential Unforgeability under Chosen Message Attack (EU-CMA) model \cite{guo2018notions}. The EU-CMA experiment for an \pkosgn~signature scheme is defined as follows:




\begin{definition} \label{def:EUCMA}
	The EU-CMA experiment $\mathit{Expt}_{\pkosgn}^{\eucma}$ for an \pkosgn~scheme is defined as follows:  
	
	\begin{itemize}[leftmargin=*]
		\item [-] $(\pk, \sk, I) \as \pkosgnkg(1^\kappa, \overrightarrow{ID})$
		\item [-] $(M^*, \sigma^*) \as \mathcal{A}^{\pkosgnsig_{\sk}(.),~\pkosgnpkconstr(.)} (PK)$:
		\item [-] If $1\as\pkosgnver(\pk, M^*, \sigma^*)$ and $M^*$ was not queried to $\pkosgnsig_{\sk}(.)$, then return $1$, else $0$.
	\end{itemize}
	
	The advantage of \A~in this experiment is defined as $ \advsaleh(\mathcal{A})= Pr[\mathit{Expt}_{\pkosgn}^{\eucma} = 1]$. The \eucma~advantage of \pkosgn~is defined as $ \advsaleh(t, q_s) = \underset{\mathcal{A}}{\textnormal{max}}\{\advsaleh(\mathcal{A})\}$, where $t$ is the time complexity of \A and $q_s$ is the number of queries to the public key constructor and signing oracles. $\pkosgnsig_{\sk}(.)$ and $\pkosgnpkconstr(.)$ are as follows:
	
    \begin{enumerate}[leftmargin=*]
        \item \textit{Signing oracle $\pkosgnsig_{\sk}(.)$}: 	Given an input message $M$, \sloppy it outputs a signature $\sigma \as \pkosgnsig_{\sk}(M)$.
        \item Public key construct oracle $\pkosgnpkconstr(.)$: Given the public key \pk, user identity $ID_i$, and counter $j$, it returns the one-time public key $\pk_i^j$. Note that unlike previous public key constructors (e.g., \cite{Yavuz:HASES:ICC2023, behnia2021towards}), $\pkosgnpkconstr(.)$~does not require a root of trust on introduced entities (e.g., \cite{behnia2021towards, nouma2024trustworthy}) and can be run based on public key data. $\pkosgnpkconstr(.)$ may be run by the verifier or a resourceful third party. 
    \end{enumerate}
 
	
	%
	%
	%
	%
\end{definition}

\section{The Proposed Scheme}
\label{sec:ProposedSchemes}
We first present our proposed scheme, \infhors. We then describe its instantiations, design rationale, and optimizations. 
%
The main bottleneck of hash-based digital signatures is the generation and management of one-time public keys. As outlined in Section \ref{subsec:Relatedwork}, the existing alternatives rely on hyper-tree structures that incur extreme signature generation and transmission overhead. A trivial yet insecure approach would be to share the master secret key with a trusted party that replenishes one-time keys for the verifiers (e.g., \cite{ouyang2021scb}). However, this invalidates the non-repudiation and makes the system vulnerable to key compromises. Moreover, it is not scalable to large-IoTs due to the massive transmission overhead. 

We address this public key management conundrum by introducing a novel approach that permits verifiers to construct one-time keys from a master public key via encrypted evaluations. Our idea is to wrap the master secret key with homomorphic encryption and then enable any verifier to retrieve one-time public keys for any valid signer $ID_i$ and message $M_i^j$. This allows signers to achieve optimal efficiency concerning $\hors$ since it only computes and broadcasts one $\hors$ signature per message. The verifiers can construct one-time public keys via encrypted evaluations without the risk of private key compromises. Our approach effectively transforms one-time $\hors$ into practically unbounded hash-based signature with minimal signer computations, and therefore, fittingly, we name our new scheme \ensuremath{\texttt{LiteQSign}}. We provide the details of $\infhors$ in Figure \ref{alg:infhors}.


\begin{figure}
	\centering
	\begin{minipage}{0.74\textwidth}
		\centering
		\noindent \fbox{\parbox{\columnwidth} {
				\scriptsize
                
	\begin{algorithmic}[1]
		
		\Statex $\underline{(\pk,  \overrightarrow{\gamma}, I)\as \infhorskg(1^{\kappa}, \overrightarrow{ID} = \{ID_i\}_{i=1}^{N})}$: 
		\vspace{2pt}
		
		\State $\msk\Ra\{0,1\}^{\kappa}$ and set $I \as (I_\horsss=(k, t), I_\fhe, I_\DM)$ according to Definitions \ref{def:HORS}, \ref{def:FHE}, \ref{def:DM}.
		\For{$i = 1, \ldots,  N$}
		\State $\gamma_i \as \prf(\msk, ID_i)$
		\EndFor
		\State $(\sk', \pk', I_{\fhe}) \as \fhekg (1^\kappa)$
		\State $\mpk \as \fheenc(\pk', \msk)$, $\pk=\langle \pk', \mpk\rangle$
		\State \Return ($\pk, \overrightarrow{\gamma}=\{\gamma_i\}_{i=1}^N, I$), where $\gamma_i$ is securely given to $ID_i$
	\end{algorithmic}
	\algrule
	\begin{algorithmic}[1]
		\Statex $\underline{\sigma_{i}^{j} \as \infhorssig(\gamma_i, M_i^j)}$: The signer $ID_i$ computes a signature on a message $M_i^j$ as follows:
		\vspace{2pt}
		\State $\sk_{i}^j \as \prf(\gamma_i, j)$
		\State $h_i^j \as H(M_i^j)$, split $h_i^j$ into $k$ sub-strings $\{h_i^{j,\ell}\}_{\ell =1}^{k}$ where $|h_i^{j,\ell}|=\log_2 t$, and interpret each $\{h_i^{j,\ell}\}_{\ell =1}^{k}$ as an integer $\{x_i^{j,\ell}\}_{\ell =1}^{k}$.
		\For{$\ell = 1, \ldots, k$}
		\State	$s_{i}^{j,\ell} \as \prf(\sk_i^j, x_i^{j,\ell})$
		\EndFor 
		\State Set $j \as j + 1$ 
		\State \Return $\sigma_i^j = (s_{i}^{j,1}, s_{i}^{j,2}, ..., s_{i}^{j,k}, j)$  
	\end{algorithmic}
	\algrule
	
	\begin{tcolorbox}[colframe=black, colback=gray!10, boxrule=0.4mm, sharp corners, boxsep=0pt, left=1mm, right=1.5mm]
    \begin{algorithmic}[1]
		\Statex   $\underline{cv_i^j \as \infhorspkconstr(\pk, ID_i, j)}$: Performed by the verifier for a given $ID_i\in \overrightarrow{ID}$ and state $j$, in {\em offline}  mode before receiving signatures, or optionally outsourced to a powerful entity.		
		\State $c\gamma_i \as \fheeval(\pk', \prf(\mpk, ID_i))$
		\State $\csk_i^j \as \fheeval(\pk', \prf(c\gamma_i, j))$
		\For{$\ell = 1, \ldots, t$}
		\State $cv_i^{j,\ell} \as \fheeval(\pk', f(\prf(\csk_i^j, \ell)))$
		\EndFor
		\State \Return $cv_i^{j} \as (cv_i^{j,1}, cv_i^{j,2}, ..., cv_i^{j,t})$
	\end{algorithmic}
    \end{tcolorbox}
	\begin{algorithmic}[1]
		\Statex $\underline{b_i^j \as \infhorsver(\pk, M_i^j,\sigma_i^j)}$:
		\vspace{2pt}
		\State $cv_i^j \as \infhorspkconstr(\pk, ID_i, j)$ \Comment{can be executed in offline mode}
		\State Execute Step 2 in \infhorssig
		\For{$\ell = 1, \ldots, k$} 
		\State $v_i^{j,\ell} \as f(s_i^{j,\ell})$
		\State $CV_\ell^j \as \fheenc(\pk', v_\ell^j)$
		\State $b_i^{j,\ell} \as \fheeval(\pk', \cmp(cv_i^{j,x_i^{j,\ell}}, CV_i^{j,\ell}))$
		\EndFor
		\If{$b_i^{j,\ell} = 1, \forall \ell=1,\ldots,k$}, \Return $b_i^j = 1$  \textbf{else}, \Return $b_i^j = 0$
		\EndIf
	\end{algorithmic}

}}
		
	\end{minipage}
\caption{LiteQSign Scheme (\infhors)}\label{alg:infhors}
	\vspace*{-2mm}
\end{figure}

The key generation algorithm $\infhorskg$, first derives the master signing key $\msk$ and sets up the public parameters $I$ including $\hors$, $\fhe$, and $\DM$ parameters as in Definition \ref{def:HORS}, \ref{def:FHE}, \ref{def:DM}, respectively (Step 1). It derives the initial private key $\gamma_i$ (seed) of each signer $ID_i$ (Step 2-3).  It then generates an $\fhe$ key pair $(\sk', \pk')$, encrypts $\msk$ with $\pk'$ to generate the master public key $\mpk$, and sets $\infhors$ public key as $PK=(\pk',\mpk)$ (Step 5). As elaborated in public key construction,  this permits any verifier to extract a one-time public key from the master public key under encryption without exposing it.  Finally, all key pairs are distributed to the verifiers and signers (Step 6).

The signature generation $\infhorssig$ for signer $ID_i$ begins by deriving the private key $\sk_{i}^{j}$ from the seed $\gamma_i$ based on the message state (counter) $j$ (Step 1). The signing process follows $\horssig$, except that the signature elements $\{s_i^{j,\ell}\}_{\ell=1}^k$ are computed via $\prf$ evaluations using $\sk_{i}^{j}$ instead of random generation (Steps 2-4). Finally, the signer updates the state $j$ and discloses the $\hors$ signature (Step 5).


$\infhorspkconstr$ algorithm enables any verifier to generate the one-time public key $\pk_i^j$ under $\fhe$ encryption associated with a valid  $ID_i\in \overrightarrow{ID}$ without any interaction with the signer or having to access private keys ($\msk, \sk'$). It first derives the initial seed $\gamma_i$ of $ID_i$ under $\fhe$ encryption that is preserved in $c\gamma_{i}$ (Step 1). It then pinpoints the private key $\sk_{i}^{j}$ of state $j$, which is sealed under $\csk_{i}^{j}$ (Step 2). Note that the signer used $\sk_{i}^{j}$ to derive $\hors$ signature components for $M_{i}^{j}$. Finally, it generates the $\fhe$ encryption of $\hors$ one-time public key for the state $j$ by evaluating $f(.)$ and $\prf$ under encryption (Step 4-5).

The signature verification $\infhorsver$ resembles $\horsver$, but starts by constructing public keys using $\infhorspkconstr$ and the signature verification is performed under encryption. The verifier performs $f$ evaluations on the received $k$ elements of the signature subset and encrypts the output using $\fhe$. Next, the verifier evaluates the comparison function $\cmp$ under encryption via $\fheeval$. As we will shortly discuss in Section \ref{subsec:SchemeInstantiationOpt}, the verifier may construct public keys offline before receiving the message-signature pair. Additionally, to reduce the storage demands, the verifier may use an alternative method by providing the indices (i.e., $\{x_i^{j,\ell}\}_{\ell=1}^k$ in Step 2, $\infhorssig$) instead of the counter $j$ to the $\infhorspkconstr$ routine. 


\subsection{\infhors~Instantiations and Optimizations} \label{subsec:SchemeInstantiationOpt}
The generic $\infhors$ in Algorithm \ref{alg:infhors} can be instantiated with any $\fhe$, $\prf$ and $f(.)$ as $\owf$. However, these instantiation choices make a drastic impact on performance, security, and practicality. In the following, we articulate our instantiation rationale and their potential optimizations. 

{\em BGV Cryptosystem as the \fhe~Instantiation}: There exist various classes and schemes of FHE. We instantiated our $\fhe$ with BGV cryptosystem \cite{brakerski2014leveled} for the following reasons:
(i) BGV is considered as a benchmark for $\fhe$ instantiations. It is well-studied and implemented in different libraries like HElib.   
(ii) We employ the Ring-Learning With Error (R-LWE) based BGV that offers an ideal security-efficiency trade-off.
(iii) BGV is amenable to parallelism and supports CRT-based encoding techniques to allow entry-wise arithmetic. (iv) It facilitates leveled-FHE, enabling the evaluation of a predetermined depth circuit without necessitating any bootstrapping.

{\em Performance Hurdles of Traditional Cryptographic Hash Functions in $\fhe$ Settings:} Presuming it takes hundreds of clock cycles for a modern processor to handle a single block cipher encryption, it takes millions of clock cycles to complete the same task under $\fhe$. Since $\infhorspkconstr$ requires $\fhe$ evaluations, we require FHE-friendly cryptographic primitives that suit the needs of $\infhors$.  
The hash-based signatures usually rely on traditional hash functions $H$ to realize both the message compression and one-way function $f(.)$. However, it was shown that ARX-based primitives like SHA-256 and BLAKE are not suitable for $\fhe$ evaluations. For instance, SHA-256 requires $3311$ $\fhe$ levels, which is infeasible for many practical purposes \cite{10.1007/978-3-642-45239-0_3}. Recent efforts have explored homomorphic evaluation of hash functions such as SHA256, SM3, etc., utilizing FHE schemes like TFHE \cite{chillotti2020tfhe} that enable rapid bootstrapping. However, they remain considerably distant from practical application, with execution times on the order of minutes \cite{bendoukha2022practical, wei2023fregata}.

{\em Mitigating Encrypted Evaluation Hurdles via Davies-Meyer as $\owf$:} We made a key observation that $f(.)$ needs only $\owf$ property but not a full cryptographic hash function. This permits us to consider alternative hash designs that rely on symmetric ciphers that are suitable for $\fhe$ evaluations. Consequently, we can leverage the best properties from both cryptographic realms.

The symmetric ciphers generally have lower multiplicative complexity (depth and size) compared to the traditional hash functions, with cheaper linear operations favoring more efficient $\fhe$ evaluations. Moreover, when evaluated under encryption, they can serve as $\owf$ with proper instantiations. We have investigated various options and identified that a block cipher-based hash function named, Davies-Meyer (DM) \cite{Preneel2005DaviesMeyerHF}, satisfies our efficiency and $\owf$ prerequisites for the encrypted evaluation purposes. Compared to other constructions, $\DM$ structure is lighter than one-way double-block-length compression methods (e.g. Hirose \cite{10.1007/11799313}), and allows for key-setup and encryption parallelization as opposed to other single-block-length one-way compression functions.

{\em Selection of Suitable Cipher for $\DM$ Instantiaton:}  
We adopt AES as the underlying primitive for our $\DM$ instantiation based on several compelling factors. (i) AES is a widely adopted and standardized block cipher with numerous optimized implementations ranging from high-performance commodity platforms to resource-constrained embedded MCUs. (ii) It features a low number of rounds and avoids complex integer operations, making it suitable for constrained environments. (iii) The algebraic structure of AES is particularly amenable to parallel computation, batching via packing techniques, and hardware-level acceleration such as GPU-based processing \cite{10.1007/978-3-642-45239-0_3}. (iv) In comparison to hash-based alternatives, an AES-based $\DM$ requires a smaller, fixed-size memory footprint for iteratively storing intermediate hash values, which is advantageous for memory-constrained deployments. (v) Finally, homomorphic evaluation of AES has been extensively studied and is supported by established libraries (e.g., HElib \cite{halevi2020design}), further affirming its suitability for our design.

{\em Optimizations:} To enhance the efficiency of signature verification, we incorporate a suite of online-offline optimizations that strategically decouple costly computations from time-sensitive operations. These optimizations shift the computational burden to the offline phase, thereby enabling faster verification during the online phase. 
\begin{itemize}[leftmargin=*]
    \item[{\em (i)}] The construction of the public key is independent of the message being verified and can be performed in advance for any identity $ID_i$ and corresponding state information. This allows the verifier to execute $\infhorspkconstr$ in batch mode offline and precompute the encrypted public keys. These precomputed values can then be efficiently leveraged for rapid online verification. As demonstrated in Section~\ref{sec:performance}, this design yields substantial performance improvements for the online phase. 
    \item[{\em (ii)}] Instead of computing the entire set of $t$ one-time public keys, the verifier may selectively generate only the $k$ components required for a given verification. This selective construction significantly reduces the number of $\fhe$ evaluations required, thereby lowering both computation time and resource usage. 
    \item[{\em (iii)}] Since $\infhorspkconstr$ is publicly executable and does not require access to any private input, the verifier may optionally delegate its offline execution to a more resource-capable external entity, such as a cloud server. In return for tolerating minor transmission latency, the verifier offloads the intensive $\fhe$ evaluations, allowing the delegated server to exploit extensive parallelism and GPU acceleration. This is particularly beneficial in the context of our $\infhors$ instantiations, which are designed to take advantage of such computational enhancements.
\end{itemize}


\section{Security Analysis} \label{sec:security}
We prove that \infhors~is EU-CMA secure as follows.

\begin{theorem} \label{the:Inf-HORSSecurityProof} 		${\adveucmaINFHORS(t, q_s)  \le  q_s \cdot \adveucmaHORS(t', q'_s)}$, where ${q’_s = q_s + 1}$ and $\mathcal{O}(t’) = \mathcal{O}(t) + q_s\cdot(k\cdot\prf + (t+2)\cdot \fheeval(\prf))$ (we omit terms negligible in terms of $\kappa$).
\end{theorem}

\noindent {\em Proof:} Let \A be the $\infhors$ attacker. We construct a  simulator \F~that uses \A as a subroutine to break one-time EU-CMA secure \hors, where ${(\overline{\sk}, \overline{\pk}, I_{\horsss}) \as \horskg(1^\kappa)}$ (Definition \ref{def:HORS}).  \F is given the challenge $\overline{\pk}$, on which \A aims to produce a forgery. \F has access to the $\horsss$ signing oracle under secret key $\overline{\sk}$. \F maintains two lists \lm~and \ls~to record the queried messages and ${\infhorssig_{\sk}(.)}$ outputs. \F randomly chooses a target forgery index\footnote{We follow SPHNICS+ \cite{cooper2024stateless} where the maximum number of signing queries is ${2^{40} \le q_s \le 2^{60} \ll 2^\kappa}$} ${w \in [1, q_s]}$. \A uses a user identity ${ID_i \in \overrightarrow{ID}}$, where ${i \Ra \{1,\ldots,N\}}$. \vspace{2mm} 


\hspace{-5mm} \textit{\noindent \underline{{\em Algorithm \F($\overline{\pk},I_{\horsss}$)}}} \vspace{2mm} 

$\bullet$~\underline{{\em Setup:}} \F is run as in Definition \ref{def:EUCMA}:

\noindent(1) ${\msk\Ra\{0,1\}^{\kappa}}$. 

\noindent(2) ${I\as(I_\horsss, I_\fhe, I_\DM)}$, where $(I_\fhe, I_\DM)$ are as in Definition \ref{def:FHE}-\ref{def:DM}, respectively.  

\noindent(3)  ${(\sk', \pk', I_{\fhe}) \as \fhekg (1^\kappa)}$. 

\noindent(4) $\mpk \as \fheenc(\pk', \msk)$; $PK = (\pk',\mpk)$.


\noindent(5)  ${\sk_i^0 \as \prf(\msk, ID_i)}$.

\noindent(6) ${sk=\{\sk_{i}^j \as \prf(\sk_i^0, j)\}_{j=1, j \neq w}^{q_s}}$. 

\noindent(7) ${\{cv_i^j \as \infhorspkconstr(\pk, ID_i, j)\}_{j=1, j \neq w}^{q_s}}$.  \vspace{2mm}

\textit{\vspace{2mm} \underline{{\em Execute $\mathcal{A}^{\infhorssig_{\sk}(\cdot),~\infhorspkconstr(\cdot),~\horssig_{\overline{\sk}}(.)} (\pk, \overline{\pk})$} }}: 

$\bullet$~\underline{{\em Queries:}} \F handles \A's queries as follows:

{\em (1) $\infhorssig_{\sk}(.)$:} \F returns $\sigma_i^w \as \horssig_{\overline{\sk}}(M_i^w)$ by querying $\hors$ signing oracle, if $j=w$. Otherwise, \F runs  the steps (2-5) in $\infhorssig$ to compute $\sigma_i^j$ under $\sk_i^j $. \F inserts $M_i^j$ to \lm and $(M_i^j, \sigma_i^j)$ to \ls as $\sigma_i^j \as \ls[M_i^j]$.


\vspace{2mm}	
{\em (2) ${\infhorspkconstr(.)}$ Queries:} If ${j=w}$ then \F returns ${cv_i^w = \fheenc(PK', \overline{\pk}}$). Otherwise, \F returns $cv_i^j$. 

\vspace{2mm}
$\bullet$~\underline{{\em Forgery of \A}}: \sloppy \A produces a forgery ${(M^*, \sigma^*)}$ on $\pk$. \A wins the $\eucma$ experiment if ${1 \as \infhorsver(\pk, M^*,\sigma^*)}$ and ${M^* \notin \lm}$ conditions hold, and returns 1, else returns $0$.

\vspace{2mm}
$\bullet$ \underline{{\em Forgery of \F:}}\vspace{+0.5mm} If \A fails to win the $\eucma$ experiment for $\infhors$, \F also fails to win the $\eucma$ experiment for $\horsss$. As a result, \F \textit{aborts} and returns $0$. Otherwise, \F checks if ${1 \as \horsver(\overline{PK},M^*,\sigma^{*})}$ and $M^*$ was not queried to the $\hors$ signing oracle (i.e., ${\horssig_{\overline{\sk}}(.)}$). If these conditions hold, \F wins the $\eucma$ experiment against $\hors$ and returns 1. Otherwise, \F \textit{aborts} and returns $0$.

\vspace{+1mm}
$\bullet$ \underline{{\em Success Probability Analysis:}}\vspace{+0.5mm} We analyze the events that are needed for \F to win the $\eucma$ experiment as follows:

\vspace{+1mm}
{\em (1) \F does not \textit{abort} during \A's queries with $Pr[\overline{Abort1}]$}:  \F can answer all of \A's signature queries, since it knows all private keys except $j=w$, for which it can retrieve the answer from $\hors$ signature oracle. \F sets $\pk_i^w=\fheenc(PK', \overline{PK})$ and can answer all other queries by running the public key construction algorithm. The only exception occurs if  $\fheeval(.)$ produces an incorrect $\pk_i^j$ during the simulation, which occurs with a negligible probability in terms of $\kappa$ due to the correctness property of $\fhe$. Therefore, we conclude $Pr[\overline{Abort~1}] \approx 1$.

\vspace{+1mm}
{\em (2) \A produces a valid forgery with $Pr[Forge|\overline{Abort1}]$}: If \F does not abort during the queries, then \A also does not abort, since its simulated view is computationally indistinguishable from the real view (see indistinguishability argument below).  Hence, the probability that \A produces a forgery against $\infhors$ is $Pr[Forge|\overline{Abort1}]=\adveucmaINFHORS(q_s, t)$. There are three events that may also lead to \A's forgery: (i) \A breaks the subset-resiliency of $H$, whose probability is negligible in terms of $\kappa$~\cite{reyzin2002better}. (ii) \A breaks IND-CPA secure FHE and recovers the master secret key $\msk$, which permits a universal forgery. The probability that this happens is negligible in terms of $\kappa$ for sufficiently large security parameters~\cite{brakerski2014leveled}. (iii) \A breaks the evaluation of the comparison circuit for all $k$ signatures (i.e., $b_i^{j,\ell} = 1, \forall \ell = 1,\ldots,k$), which occurs with a probability that is $\frac{1}{k} \times$ negligible in relation to $\kappa$. (iv) \A inverts $\DM$ by breaking the underlying IND-CPA cipher, which also happens with negligible probability in terms of $\kappa$ \cite{Preneel2005DaviesMeyerHF}. Therefore, they are omitted in the theorem statement. 

\vspace{+1mm}
{\em (3) \F does not \textit{abort} after \A's forgery with $Pr[\overline{Abort2}|\overline{Abort1} \wedge Forge]$}: \F does not abort if \A's forgery is on the target public key $PK_i^w$. Since $w$ is randomly selected from $[1,q_s]$, this occurs with $1/q_s$. 

\vspace{+1mm}
{\em (4) \F wins the $\eucma$ experiment with $\adveucmaHORS(t', q'_s) $}: 

$Pr[Win]= Pr[\overline{Abort 1}]$ $\cdot Pr[Forge|\overline{Abort1}] \cdot Pr[\overline{Abort2}$ $|\overline{Abort1} \wedge Forge]$. Therefore, $Pr[Win]$ is bounded as: 
\vspace{-0.3mm}
\begin{eqnarray*}
	\adveucmaINFHORS(t, q_s) & \le & q_s \cdot \adveucmaHORS(t', q'_s) 
\end{eqnarray*}

$\bullet$~\underline{{\em Execution Time Analysis:}}\vspace{+0.5mm} The running time of \F is that of \A plus the time required to respond to $q_s$ public key and signature queries. Each signature query demands $H$ and $k\cdot \prf(.)$; and each $\infhorspkconstr(.)$ query needs $(t+2)\cdot\fheeval(\prf)$ . The approximate running time of \F is $\mathcal{O}(t') = \mathcal{O}(t) + q_s\cdot(k\cdot\prf + (t+2)\fheeval(\prf))$. \vspace{2mm}

$\bullet$~\underline{{\em Indistinguishability Argument}}:\vspace{+0.5mm} In the real view of \A ($\mathcal{A}_{real}$), all values are computed from the master secret key and seeds as in the key generation, signing, and public key construction algorithms. The simulated view of \A ($\mathcal{A}_{sim}$) is identical to $\mathcal{A}_{real}$, except $\pk_i^w$ is replaced with the challenge $\hors$ public key. This implies that  $(\sk_i^w=\overline{\sk}, \pk_i^w=\overline{\pk})$ holds.  Since $\horskg(.)$ generates the secret keys random uniformly (Definition \ref{def:HORS}), the joint probability distribution of  $(\sk_i^w, \pk_i^w)$ in  $\mathcal{A}_{sim}$ is similar to that of $\mathcal{A}_{real}$. Therefore, $\mathcal{A}_{real}$ and $\mathcal{A}_{sim}$ are computationally indistinguishable. $\blacksquare$

\begin{mycorollary} \label{cor:PQSecurity} 		
	The $\infhors$ scheme provides PQ promises.
\end{mycorollary}

\noindent\textit{Proof}: Based on our preceding formal security analysis and the incorporation of cryptographic primitives such as FHE, PRF, and hash functions, the $\infhors$ scheme ensures PQ assurances. Specifically, the PRF and hash functions, being symmetric cryptography primitives, remain unaffected by Shor's algorithm, while the impact of Grover's probabilistic algorithm can be mitigated by scaling up the sizes, considering the potential of quantum computers. Additionally, the FHE schemes, exemplified by our instantiation, the BGV scheme \cite{brakerski2014leveled}, are constructed upon lattice-based hard problems (e.g., General-LWE), which provide PQ security. $\blacksquare$

\section{Performance Analysis and Comparison} \label{sec:performance}
In this section, we give a detailed performance analysis of \infhors~and compare with its counterparts.

\subsection{Evaluation Metrics and Experimental Setup}

\noindent \underline{\textit{Evaluation Metrics:}} 
Our analysis evaluates $\infhors$ and its analogous counterparts, with a main focus on the signer efficiency that includes:
\textbf{(i)} private key and signature sizes which translates into small memory footprint and low memory access requirements. This not only reduces the energy consumption but also frees up more memory for main applications. It is particularly important for low-end IoT devices, which are characterized by limited memory space and relatively expensive memory access (e.g., 8-bit AVR microcontrollers). 
\textbf{(ii)} signing computational efficiency which translates into reduced energy consumption and longer battery lifetime for resource-limited devices. 
\textbf{(iii)} long-term security (i.e., PQ security) in order to offer resiliency against the quantum computing breaches (e.g., Shor's algorithm \cite{shor1999polynomial}). 

\vspace*{2mm}
\noindent \underline{\textit{Parameter Selection}}:
Our system-wide parameters are $I=(I_\hors, I_\fhe, I_\DM)$. We choose $I_{\hors} \as (k=16,t=1024)$, where SHA-256  and $\DM$ are used as $H$ and $f$ (i.e., $\owf$), respectively. In $I_\DM$,  we choose AES-128 as our $\prf$.  In $I_\fhe$, we set the plaintext space of mod $2$, the lattice dimension $\phi(m) = 46080$, where the $m$-th cyclotomic is $m = 53261$. We utilize a packing technique that empowers us to evaluate $120$ blocks of AES at once. We set $N=2^{20}$ as the number of resource-constrained signers within the IoT network. 
\vspace*{2mm}

\noindent \underline{\textit{Hardware Configuration}}:
We tested $\infhors$ on both commodity hardware and a low-end MCU.

\noindent $\bullet$ {\em Commodity Hardware:} is a resourceful desktop equipped with an Intel i9-9900K@3.6GHz processor and 64GB of RAM.

\noindent  $\bullet$ {\em Embedded device:} We evaluate $\infhors$ on an 8-bit ATxmega128A1 microcontroller to assess its efficiency on embedded IoT devices. The microcontroller features 128 KB flash memory, 2 KB RAM, 8 KB EEPROM, and operates at a 32 MHz clock frequency.



\vspace*{2mm}
\noindent \underline{\textit{Software Configuration}}:
For the commodity hardware, we utilized the following libraries
(i) OpenSSL\footnote{\url{https://github.com/openssl/openssl}} to implement SHA-256.
(ii) HElib\footnote{\url{https://github.com/homenc/HElib}} to implement $\fhe$ functionalities (\textit{e.g.}, evaluation and comparison under encryption\footnote{\url{https://github.com/iliailia/comparison-circuit-over-fq/tree/master}}).
(iii) $\DM$ is implemented using the hardware-optimized AES-NI \cite{hofemeier2012introduction}.
For the 8-bit AVR device, we employed the AVR cryptographic library\footnote{\url{https://github.com/cantora/avr-crypto-lib}} to implement AES-128, offering an optimized assembly implementation, resulting in mimimal cycles to evaluate hashing and $\prf$ calls. 

\vspace*{2mm}
\noindent \underline{\em Selection Rationale of Counteparts}:
The selection of our counterparts is based on the discussed evaluation metrics and the availability of open-source implementation and/or open-access benchmarks. Numerous digital signatures have been proposed in the literature that address the resource limitations of IoT devices. Nevertheless, few schemes address low-end embedded devices, such as our target 8-bit AVR MCU. In order to cover different signatures with the knowingly existing post-quantum intractability assumptions, we carefully selected the following constructions:
\begin{itemize}[leftmargin=*]
    \item[{\em(i)}] {\em lattice-based}: the NIST PQC standards ML-DSA-\romannum{2} \cite{ducas2018crystals} and Falcon-512 \cite{fouque2018falcon}. They are considered the most prominent lattice-based signatures, with balanced efficiency between key sizes and signing efficiency. We also selected BLISS-\romannum{1} because it is the only lattice-based signature with a benchmark on an 8-bit AVR MCU \cite{poppelmann2015high}. We also selected the authentication scheme ANT-II \cite{behnia2021towards}, which relies on distributed non-colluding servers to supply one-time public keys to verifiers. ANT-II is susceptible to networks delays and outages, and prone to malicious server behaviors. 
    \item[{\em(ii)}] {\em hash-based}: generally suffer from an expensive signing cost with larger key sizes. We selected the NIST PQC standard SLH-DSA \cite{cooper2024stateless}, a stateless signature scheme. We also selected $\xmss$ \cite{hulsing2013optimal} as a standard stateful hash-based signature with forward security. To our knowledge, there is no hash-based signature with a benchmark on 8-bit AVR MCUs.  
    \item[{\em(iii)}] {\em multivariate-based}: are known to be computationally efficient in terms of signing and verification with small signature and public key sizes. However, they generally suffer from large private key sizes, resulting in high memory usage and frequent access. This limitation might be problematic when deployed on highly constrained 8-bit devices with 128KB of static flash memory. There exist numerous multivariate-based digital signatures that have been proposed (e.g., \cite{li2024efficient, shim2020high}). We identified $\himq$ \cite{shim2020high} that achieves a high signing efficiency on an 8-bit AVR ATxmega384C3. However, we observed a high memory usage that includes the private key size and code size, occupying $72.38\%$ of the flash read-only memory of our target ATxmega128A1. Therefore, we omit it from our performance analysis due to the high memory usage. 
    \item[{\em(iv)}] {\em conventional signatures}: We also considered non-PQ signature schemes. Although they do not achieve long-term security, ECC-based signature schemes are signer-efficient with small key sizes. We selected the mostly-used standards, ECDSA \cite{johnson2001elliptic} and Ed25519 \cite{bernstein2012high}. Other conventional (e.g., pairing-based \cite{boneh2001short}) digital signatures incur expensive operations during signature generation, therefore, not practical for resource-limited IoT devices. 
\end{itemize}

\subsection{Performance on Signer}
Performance comparisons on commodity hardware and the embedded device are shown in Tables \ref{tab:commodityperformance} and \ref{tab:embeddedperformance}, respectively.

\begin{table*}[t!]
	\centering
	\Large
	\caption{Performance comparison of \infhors~and its counterparts on commodity hardware}
	\label{tab:commodityperformance}

	\resizebox{\textwidth}{!}{
	\begin{tabular}{|l|c|c|c|c|c|}
		\hline
		\textbf{Scheme}
		& \textbf{Signing Time ($\mu s$)} 
		& \textbf{Private Key ($KB$)} 
		& \textbf{Signature Size ($KB$)} 
		& \multicolumn{2}{c|}{\textbf{Verification Time ($\mu s$)}} 
		\\ \hline

		ECDSA~\cite{johnson2001elliptic} 
		& $16.98$ & $0.06$ & $0.06$ & \multicolumn{2}{c|}{$46.41$} 
		\\ \hline

		Ed25519~\cite{bernstein2012high} 
		& $16.39$ & $0.06$ & $0.06$ & \multicolumn{2}{c|}{$39.75$} 
		\\ \hline

		BLISS-\romannum{1}~\cite{ducas2013lattice} 
		& $244.97$ & $2.00$ & $5.6$ & \multicolumn{2}{c|}{$25.21$} 
		\\ \hline

		ML-DSA-\romannum{2}~\cite{ducas2018crystals} 
		& $93.76$ & $2.29$ & $2.36$ & \multicolumn{2}{c|}{$18.73$} 
		\\ \hline

		Falcon-512~\cite{fouque2018falcon} 
		& $184.74$ & $1.29$ & $0.65$ & \multicolumn{2}{c|}{$32.16$} 
		\\ \hline

		SLH-DSA~\cite{cooper2024stateless} 
		& $5,441.58$ & $0.13$ & $32.63$ & \multicolumn{2}{c|}{$549.63$} 
		\\ \hline

		\xmss~\cite{hulsing2013optimal} 
		& $10,682.35$ & $3.11$ & $2.61$ & \multicolumn{2}{c|}{$2,098.84$} 
		\\ \hline

		\multirow{2}{*}{\textbf{LiteQSign}} 
		& \multirow{2}{*}{\textbf{4.81}} 
		& \multirow{2}{*}{\textbf{0.02}} 
		& \multirow{2}{*}{\textbf{0.25}} 
		& \textbf{Online} & \textbf{PKConstr (Offline)}	\\ \cline{5-6}
            & & & & \textbf{1.91 s} & \textbf{41.22 s} \\ \hline
	\end{tabular}
	}

	\vspace{0.3em}

	\resizebox{\textwidth}{!}{
	\begin{tabular}{|l|c|c|c|c|c|c|}
		\hline
		\multirow{2}{*}{\textbf{Scheme}} 
		&  \multicolumn{2}{c|}{\textbf{Verifier Storage}} 
		& \textbf{Total Storage} 
		& \textbf{Post-Quantum}  
            & \textbf{Sampling}
		& \textbf{Simple} \\ \cline{2-3}
        &\textbf{Public Key ($KB$)} 
		& \textbf{Certificate ($KB$)} &\textbf{($2^{20}$ users) ($GB$)}&\textbf{Promise}&\textbf{Operations}&\textbf{Code Base}
		\\ \hline

		ECDSA~\cite{johnson2001elliptic} 
		& $0.09$ & $0.06$ & $0.16$ & $\times$ & $\times$ & $\times$ 
		\\ \hline

		Ed25519~\cite{bernstein2012high} 
		& $0.09$ & $0.06$ & $0.16$ & $\times$ & $\times$ & $\times$ 
		\\ \hline

		BLISS-\romannum{1}~\cite{ducas2013lattice} 
		& $7.00$ & $5.6$ & $12.6$ & $\checkmark$ & $\checkmark$ & $\times$ 
		\\ \hline

		ML-DSA-\romannum{2}~\cite{ducas2018crystals} 
		& $1.28$ & $2.36$ & $3.75$ & $\checkmark$ & $\checkmark$ & $\times$ 
		\\ \hline

		Falcon-512~\cite{fouque2018falcon} 
		& $0.88$ & $0.65$ & $1.53$ & $\checkmark$ & $\checkmark$ & $\times$ 
		\\ \hline

		SLH-DSA~\cite{cooper2024stateless} 
		& $0.06$ & $32.63$ & $32.69$ & $\checkmark$ & $\times$ & $\times$ 
		\\ \hline

		\xmss~\cite{hulsing2013optimal} 
		& $0.75$ & $2.61$ & $3.36$ & $\checkmark$ & $\times$ & $\times$ 
		\\ \hline

		\textbf{LiteQSign} 
		& \multicolumn{3}{c|}{\textbf{9.42 MB}} 
		& $\checkmark$ & $\times$ & $\checkmark$ 
		\\ \hline
	\end{tabular}
	}

	\begin{tablenotes}[flushleft] \scriptsize{
		\item The private/public key, signature, and certificate sizes are in KB. $\infhors$ and NIST PQC candidates use architecture-specific optimizations (i.e., AESNI, AVX2 instructions). For $\xmss$, we choose the XMSST\_MT\_SHA$2\_20\_256$ variant. For SLH-DSA, we set  $n=256, h=63, d=9, b=12, k=29, w=16$. 
	The total verifier storage denotes the storage required to verify ($J=2^{30}$) signatures for ($N=2^{20}$) signers.
	}
	\end{tablenotes}
\end{table*}

\vspace*{2pt}
\underline{\em $\bullet$ Memory Usage:} $\infhors$ achieves the lowest memory usage by having the smallest private key size among its counterparts. For example, the private key of $\infhors$ is $3\times$ and $114\times$ smaller than that of the conventional signer-efficient Ed25519 and PQ-secure ML-DSA standards, respectively. The private key is $22\times$ smaller than that of the most efficient lattice-based counterpart, BLISS-\romannum{1} \cite{ducas2013lattice}, respectively. It is without incurring large code size and expensive costly sampling operations that may result in side-channel attacks \cite{tibouchi2021one}. 
Notably, $\infhors$ consumes less memory than its most signer-efficient and PQ-secure counterpart, $\himq$ \cite{shim2020high}, by having a significantly smaller private key size. The cryptographic storage (including the code size) of $\himq$ utilizes $72.38\%$ of the overall flash memory size of an 8-bit AVR ATxmega128A1, whereas $\infhors$ utilizes only $2.8\%$. We argue that the cryptographic data should occupy minimal space, particularly in resource-limited devices (e.g., pacemakers, medical implants). 
Indeed, the embedded devices generate system-related (e.g., log files) and application-related (e.g., sensory information) data, which may cause memory overflow, considering the high memory cryptographic usage. 
It is noteworthy that we do not assess the impact of memory access on the battery lifetime of the embedded device. We foresee a high energy usage of multivariate-based signatures compared to that of $\infhors$, considering the high cost of both memory and stack usage. 


\vspace*{2pt}
\underline{\em $\bullet$ Bandwidth Overhead:} $\infhors$ boasts a compact signature size that is $9.4\times$ and $2.6\times$ smaller than the NIST PQC standards, ML-DSA-\romannum{2} and Falcon-512, respectively. The signature size of $\infhors$ is also $22\times$ smaller than that of the most-efficient lattice-based BLISS-\romannum{1}. A small signature size results in low transmission overhead, thereby minimizing energy consumption on resource-constrained IoT devices. This reduced energy expenditure is crucial for extending the operational lifespan of devices that often operate on limited power sources.

\vspace*{2pt}
\underline{\em $\bullet$ Signature Generation:}
Table \ref{tab:commodityperformance} demonstrates that among our counterparts (i.e., conventional-secure and post-quantum), $\infhors$ exhibits the fastest signing time and the lowest signer storage overhead. It is $10\times$ and $43\times$ faster than the NIST PQC standards, ML-DSA-\romannum{2} and Falcon-512, respectively. The computational performance advantages at the signer of $\infhors$ become even more apparent on embedded devices. Based on 8-bit AVR MCU results in Table \ref{tab:embeddedperformance}, the signing time of $\infhors$ is $20\times$ and $44\times$ faster than the most efficient PQ-secure BLISS-\romannum{1} and conventional-secure Ed25519, respectively. 

\begin{table*}[ht!]
	\centering
	\caption{Performance comparison of \infhors~and counterparts at the signer on 8-bit AVR AtMega128A1 MCU} 
	\label{tab:embeddedperformance}
	\resizebox{0.9\textwidth}{!}{
	\begin{tabular}{|l|c|c|c|c|c|}
		\hline
		\multirow{2}{*}{\textbf{Scheme}} 
		& \multicolumn{2}{c|}{\textbf{Signing Overhead}} 
		& \multicolumn{2}{c|}{\textbf{Transmission Overhead}} 
		& \multirow{2}{*}{\textbf{Total Energy (mJ)}}
		\\ \cline{2-5}
		& \textbf{Time ($cycles$)} & \textbf{Cost ($mJ$)} 
		& \textbf{Signature Size ($KB$)} 
		& \textbf{Energy ($mJ$)} 
		& 
		\\ \hline

		ECDSA~\cite{johnson2001elliptic} 
		& $34{,}903{,}000$ & $1.52$ 
		& $0.06$ & $0.08$ & $1.60$ 
		\\ \hline
		
		Ed25519~\cite{bernstein2012high} 
		& $22{,}688{,}583$ & $1.13$ 
		& $0.06$ & $0.08$ & $1.21$ 
		\\ \hline
		
		BLISS-\romannum{1}~\cite{ducas2013lattice} 
		& $10{,}537{,}981$ & $0.49$ 
		& $5.6$ & $7.70$ & $8.19$ 
		\\ \hline
		
		\textbf{LiteQSign} 
		& $\boldsymbol{514{,}788}$ & $\boldsymbol{0.02}$ 
		& $\boldsymbol{0.25}$ & $\boldsymbol{0.34}$ & $\boldsymbol{0.36}$ 
		\\ \hline
	\end{tabular}
	}

    \vspace{0.4em}

	\resizebox{0.9\textwidth}{!}{
	\begin{tabular}{|l|c|c|c|c|c|c|c|c|c|}
		\hline
		\multirow{2}{*}{\textbf{Scheme}} 
		& \multicolumn{4}{c|}{\textbf{Expected Operation w.r.t. Transmission Frequency}} 
		& \multirow{2}{*}{\specialcell{\textbf{Secret} \\ \textbf{Key ($KB$)}}} 
		& \multirow{2}{*}{\textbf{PQ}} 
		& \multirow{2}{*}{\textbf{EoI}} 
		& \multirow{2}{*}{\textbf{BC}} 
		& \multirow{2}{*}{\textbf{SCA}}
		\\ \cline{2-5}
		& \textbf{f=1 $sec$} & \textbf{f=2 $sec$} & \textbf{f=10 $sec$} & \textbf{f=1 $min$}
		& & & & &
		\\ \hline

		ECDSA~\cite{johnson2001elliptic} 
		& 48.85d & 97.71d & 1.34y & 8.01y 
		& $0.06$ & $\times$ & $\times$ & $\checkmark$ & $\checkmark$
		\\ \hline

		Ed25519~\cite{bernstein2012high} 
		& 64.55d & 129.10d & 1.77y & 10.65y 
		& $0.06$ & $\times$ & $\times$ & $\checkmark$ & $\checkmark$
		\\ \hline

		BLISS-\romannum{1}~\cite{ducas2013lattice} 
		& 9.54d & 19.08d & 95.42d & 1.56y 
		& $2$ & $\checkmark$ & $\times$ & $\times$ & $\checkmark$
		\\ \hline

		\textbf{LiteQSign} 
		& \textbf{217.01d} & \textbf{1.19y} & \textbf{5.95y} & \textbf{35.68y} 
		& $\boldsymbol{0.02}$ & $\checkmark$ & $\checkmark$ & $\checkmark$ & $\times$
		\\ \hline
	\end{tabular}
	}

	\begin{tablenotes}[flushleft] 
        \scriptsize{
		\item The counterpart selection covers the most efficient existing conventional (ECDSA, Ed25519) and PQ-secure (BLISS), with an available benchmark on the selected 8-bit AVR MCU. PQ denotes post-quantum security. EoI denotes ease of implementation. BC denotes backward compatibility. SCA denotes side-channel attacks found in the literature.
	} 
	\end{tablenotes}
\end{table*}

\vspace*{2pt}
\underline{\em $\bullet$ Energy Consumption:}
The high signing efficiency translates into better energy awareness on low-end IoT devices. To demonstrate the potential of $\infhors$, in Table \ref{tab:embeddedperformance} presents a comprehensive energy analysis that measures the energy consumption per signature generation and transmission. We follow \cite{piotrowski2006public}, which considers a MICaz sensor node operating on an ATmega128L MCU, equipped with a ZigBee 2.4GHz radio chip (CC2420), and powered by an AA battery with an energy capacity of 6750J. The sensor node drains 4.07nJ per cycle and 0.168$\mu$J per bit transmission.  
We measure the energy consumption during one signature generation and transmission. We then measure the expected operation time of the IoT device based on different signing frequencies and assuming that the device performs only signature generation operations. Our findings reveal that \infhors~can operate with a signature generation and transmission frequency equal to 10 seconds and one minute for up to $5.95$ and $35.7$ years without battery replacement, respectively, while it is $95$ days and $1.56$ years for our sole PQ BLISS PQ counterpart, which is also not backward compatible with currently deployed cryptographic primitives.  Moreover, \infhors~performs better than the conventional EC standards (ECDSA,Ed25519) by drawing $4.44\times$ and $3.36\times$ less energy capacity, respectively.

Remind that, to the best of our knowledge, none of the selected NIST PQC signature standards have an open-source implementation available on resource-limited devices (i.e., 8-bit microcontrollers). The most prominent PQ alternatives with a reported performance on this platform are  $\text{HiMQ-3}^{\text{big}}$ \cite{shim2020high} and BLISS-I \cite{ducas2013lattice}.  We also included the most efficient ECC-based alternative Ed25519 and the widely-used ECDSA in our energy comparison to assess \infhors~performance with respect to (pre-quantum) conventional schemes. 
Our energy analysis showcases that $\infhors$ offer the longest battery lifetime when only the cryptographic computation is considered. Hence, we confirm that $\infhors$ is the most suitable signature scheme for highly resource-constrained IoT devices. 


We note that BLISS-\romannum{1} is vulnerable to side-channel attacks, which hinders its use in practice.
Side-channel attack resiliency and ease of implementation are important factors for the practical deployment of signature schemes on embedded devices. Lattice-based signatures require various types of sampling operations (e.g., Gaussian, rejection samplings) that make them vulnerable to side-channel attacks \cite{karabulut2021falcon}.  Moreover, due to their complexity, they are notoriously difficult to implement on such platforms. As an example, Falcon needs 53 bits of precision to implement without emulation \cite{howe2022benchmarking}, which hinders its deployment on 8-bit microcontrollers. 
$\infhors$ signature generation requires only a few $\prf$ calls. Hence, it is free from the aforementioned specialized side-channel and timing attacks that target sampling operations. Moreover, it is easy to implement since it only requires a suitable symmetric cipher (e.g., AES) and a cryptographic hash function (e.g., SHA256) with a minimal code size. 
Our analysis validates that $\infhors$ is the most suitable alternative among its counterparts to be deployed for signing on IoT applications due to its high computational efficiency, compact key and signature sizes, and high security. 

\subsection{Performance on Verifier} 
While $\infhors$ is a signer-optimal scheme, we also introduced strategies to minimize the verifier's computational and storage overhead. As explained in Section \ref{subsec:SchemeInstantiationOpt}, the verifiers can generate public keys in offline mode (before signature verification), thereby improving the efficiency of online verification. Moreover, the verifiers have the option to outsource offline public key construction to a resourceful entity.

\noindent \underline{\textit{Online Verification:}} The online verification cost is comprised of $k\times \prf(.)$, $k\times \fheenc(.)$, and $k\times \fheeval(.)$ of the comparison circuit. According to our implementation parameters, this is estimated to be approximately $1.913$ seconds. Also, for further cost reduction, we strongly recommend an offline generation of public keys whenever possible. 

In our instantiation, instead of generating a full set of $t$ keys, the verifier can only construct $k$ one-time public key components. Specifically, the verifier performs $k\times \prf(.)$ for the $k$ signature elements, $k\times \fheenc(.)$ to obtain the encrypted version of them, and $k\times \fheeval(.)$ for evaluating the comparison circuit under encryption. 

\noindent \underline{\textit{Offline Public Key Construction:}}  The principal computational overhead in $\infhors$ arises from its offline phase. Empirical evaluations indicate that a single homomorphic AES evaluation per block requires approximately $2.29$ seconds, resulting in a cumulative cost of $41.22$ seconds to generate $k$ public key components. This offline cost, however, can be substantially mitigated through parallelization strategies. Specifically, each public key element in the $\hors$ construction can be independently computed, making it well-suited for distribution across multiple threads or processing units. Additionally, as detailed in Section~\ref{subsec:SchemeInstantiationOpt}, the BGV homomorphic encryption scheme supports extensive parallelization, which was a decisive factor in selecting AES as the $\DM$ building block. 

Several FHE libraries support BGV and offer hardware-level optimizations. For instance, Microsoft SEAL\footnote{Microsoft SEAL Library \url{XXX}} and PALISADE\footnote{PALISADE Library \url{XXX}} provide support for Advanced Vector Extensions (AVX), while OpenFHE\footnote{OpenFHE Library \url{XXX}} extends this capability to include GPU acceleration. Empirical benchmarks demonstrate that OpenFHE achieves a speedup of approximately $13\times$ for multiplicative depth one, and up to $26\times$ for depth five, compared to non-accelerated baselines \cite{papadakis2024towards}.

Further efficiency gains can be achieved by leveraging the Chinese Remainder Theorem (CRT) \cite{pei1996chinese}, which facilitates the encryption of element-wise vectors, thereby enabling component-wise homomorphic operations such as additions and multiplications. This batching mechanism allows thousands of parallel function evaluations—such as AES instances—across distinct inputs. Notably, HElib has been shown in some configurations to outperform other libraries like Microsoft SEAL in batching efficiency, which is particularly relevant to our design. We intend to explore this optimization avenue in future work.


\noindent \underline{\textit{Verifier Storage Overhead:}} The total size of the master public key $\pk$ with the expansion per block evaluation is around  $9.42$ MB. If only a single signer is considered, the size of $\pk$ is much larger than that of its counterparts. However, $\infhors$ enables a verifier to construct public keys for {\em any valid} signer $ID_i$ of any state. This unique property permits $\infhors$ to achieve compact storage for a large number of signers since the verifier does not need to store a certificate for their public keys. For example, the total storage (public key plus certificate) for $2^{20}$ users is still 9.42 MB for $\infhors$, while it is around 1.52 GB and 3.74 GB for Falcon-512 and ML-DSA-II, respectively. The total storage advantage increases with a growing number of signers.

\section{Conclusion}

This work addressed the critical limitations of deploying PQ digital signatures in resource-constrained IoT environments, where signer-side efficiency is vital for energy preservation, device longevity, and reliable real-time operation. While NIST-standardized PQ signature schemes provide strong quantum security, their substantial computational, memory, and communication overhead render them impractical for low-end IoT devices such as wearables, sensors, and embedded systems. 
To overcome these challenges, we proposed \textit{LightQSign} (\infhors), a novel lightweight PQ signature scheme that achieves near-optimal signature generation through a constant number of hash operations per signing and supports non-interactive, on-demand public key reconstruction by verifiers. \infhors requires no trusted third parties, secure enclaves, or stateful key management, making it particularly suited for scalable and stateless deployment. We formally proved its security in the random oracle model and demonstrated its practical viability through comprehensive evaluations on both commodity and highly resource-constrained 8-bit AtMega128A1 microcontrollers. 
Our results show that \infhors outperforms NIST PQC standards in signer efficiency, memory footprint, and energy consumption, offering a practical, quantum-resilient authentication solution for heterogeneous IoT ecosystems. As future work, we aim to extend this architecture to support more advanced security features and explore the use of optimized and parallelized FHE techniques with GPU acceleration to reduce verifier-side overhead, further enhancing the scalability of quantum-secure systems.

\section*{Acknowledgment}
This work is supported by National Science Foundation NSF-SNSF 2444615 and Cisco Research Award (220159).

\bibliographystyle{ACM-Reference-Format}
\bibliography{ref}

\end{document}